%% file: bare_conf.tex
\documentclass[conference]{IEEEtran}
\usepackage{graphicx}
\usepackage{subcaption}
\usepackage{makecell,multirow,multicol,diagbox}
\usepackage[numbers,sort&compress]{natbib}
\usepackage{amsmath,amssymb}
\usepackage{algorithm}
\usepackage{makecell}
\usepackage{algorithmicx}
\usepackage{algpseudocode}
\usepackage{xcolor}
\newcolumntype{H}{>{\setbox0=\hbox\bgroup}c<{\egroup}@{}}
\usepackage{url}
\usepackage{todonotes}

\ifCLASSINFOpdf
\else
\fi
\begin{document}
\title{BigRoots: An Effective Approach for Root-cause Analysis of Stragglers in Big Data System}

\author{\IEEEauthorblockN{Honggang Zhou$^\dagger$, Yunchun Li$^\dagger$, Hailong Yang$^\dagger$, Jie Jia$^\diamond$, Wei Li$^\dagger$}
\IEEEauthorblockA{$^\dagger$School of Computer Science and Engineering, Beihang University, Beijing, China\\
$^\diamond$Taiyuan University of Technology, Taiyuan, China\\
Email: \{zhg, lych, hailong.yang, liw\}@buaa.edu.cn$^\dagger$, jiajie1921@link.tyut.edu.cn$^\diamond$}}
\maketitle

\begin{abstract}
Stragglers are commonly believed to have a great impact on the performance of big data system. However, the reason to cause straggler is complicated. Previous works mostly focus on straggler detection, schedule level optimization and coarse-grained cause analysis. These methods cannot provide valuable insights to help users optimize their programs. In this paper, we propose BigRoots, a general method incorporating both framework and system features for root-cause analysis of stragglers in big data system. BigRoots considers features from big data framework such as shuffle read/write bytes and JVM garbage collection time, as well as system resource utilization such as CPU, I/O and network, which is able to detect both internal and external root causes of stragglers. We verify BigRoots by injecting high resource utilization across different system components and perform case studies to analyze different workloads in Hibench. The experimental results demonstrate that BigRoots is effective to identify the root cause of stragglers and provide useful guidance for performance optimization.
\end{abstract}


\input{introduction}
\input{background}
\input{methodology}
\input{evaluation}
\input{related_work}
\input{conclusion}
\input{acknowledgement}

\bibliographystyle{IEEEtran}
\bibliography{IEEEabrv,reference}
\end{document}

%% file: introduction.tex
\section{Introduction}
\label{sec:introduction}
In the past decade, the advance of cloud computing has been unprecedentedly accumulating massive data that is beyond the storage capacity of any single machine. Popular frameworks for handling large volume of data are adopted by both companies and research institutes to extract useful information, which includes Mapreduce~\cite{mapreduce}, Dryad~\cite{dryad}, Spark~\cite{spark} and etc. The underlying method of these frameworks is to divide data into small pieces and perform calculation in parallel (these parallel calculations are called \textit{tasks}, a group of such tasks is called \textit{stage}). Only when every task within a stage finishes, the application can proceed to next stage. If certain tasks are slower than the rest in the same stage, the execution of the entire application is slowed down by these tasks (\textit{stragglers}).

Although stragglers do not occur frequently, they have great impact on the performance of big data system. Ananthanarayanan et al.~\cite{mantri} analyze trace data from Microsoft Bing and find 80\% of the stragglers have a uniform probability of delay between 150-250\% compared to median task duration and 10\% of the stragglers require 10$\times$ longer than median task duration to complete. Jeffrey et al.~\cite{straggler_google} study a real Google service and find that the slowest 5\% of the completed requests are responsible for half of the total 99\% latency. P. Garraghan et al.~\cite{straggler_impact} analyze two large-scale production systems (12,532 nodes in 29 days and 2,841 nodes in 14 days respectively) and find that 5\% of the stragglers impact 50\% of the total jobs for batch applications.

Many existing works have proposed speculative execution to mitigate the impact of stragglers. Current big data frameworks have already adopted the method of speculative execution, which launches a replicated task on another machine if a task executes much slower than others. Google claims that speculative execution improves the job response time by 44\%~\cite{google_estimation}. M. Zaharia, et al.~\cite{late} propose the LATE (Longest Approximate Time to End) speculative strategy for heterogeneous cluster. They rely on progress rate to detect possible stragglers and only launch speculative tasks on fast node with high progress rate. Ananthanarayanan et al.~\cite{dolly} focus on small jobs and introduce Dolly that clones all tasks of a small job to avoid waiting during speculation. Dolly achieves significant speedup for small jobs while consuming only 5\% extra resources. However, these speculative methods all have the same shortcoming. Speculation consumes additional resources. Production clusters usually run many jobs simultaneously, thus speculative execution will contend resources with normal jobs. E. Bortnikov et al.~\cite{speculation_waste} demonstrate that 90\% of the launched speculative tasks are unnecessary and waste resources in production environment.

There are research works trying to locate the root cause of stragglers with online analysis to improve the efficiency of speculative execution~\cite{mantri,dolly}. However, online analysis can only access limited information about tasks and cause additional overhead. In this paper, we advocate offline root-cause analysis of stragglers, which is more accurate with rich information regarding the tasks. Note that production clusters usually run the same jobs repeatedly, therefore improving the performance of such jobs with offline analysis is cost-effective. Once we identify the root causes of stragglers, we can mitigate their impact by taking corresponding optimizations. For instance, if most stragglers are due to poor data locality, the programmer should optimize the data layout.

Previous works on offline root-cause analysis fail to provide useful insight for further performance optimization in big data system, and sometimes the analysis is even mis-leading. For instance, X. Quyang et al.~\cite{cause_simple} analyze production clusters and find stragglers are correlated with high resource utilization. They conclude that there are three major causes including CPU utilization, disk utilization and slow request handling. This conclusion is mis-leading because high resource utilization can be generated by normal tasks that use resource intensively. P. Garraghan et al.~\cite{straggler_impact} use correlation and diagnosis method to identify the root causes of stragglers. However, their approach is at job level and does not provide specific reason to the cause of stragglers. Ananthanarayanan et al.~\cite{mantri} attribute the stragglers to data skew, cross rack traffic, bad and busy machines sequentially. This method is not effective since the above reasons could not cover all possible root causes of stragglers.

In this paper, we propose BigRoots, a general method incorporating features from both framework and system for root-cause analysis of stragglers, which covers a broader spectrum of causes and provides insightful guidance for performance optimization. The underlying idea of our approach is to compare the features of stragglers with normal tasks in the same stage. If the value of a straggler feature deviates greatly from that of normal task, we treat this feature as the root cause of straggler. This method overcomes the drawbacks of previous works and provides useful guidance for further performance optimization. In addition, statistical rules are applied to different features in order to reduce the false positive results of the root-cause analysis. For instance, we can filter out the features representing blocking time that are much smaller than task duration. The reason is that if the time spent on such features is insignificant, then these features would not strongly affect the performance of the task.

Specifically, this paper makes the following contributions:
\begin{itemize}
\item{We propose BigRoots, a general method for root-cause analysis of stragglers in big data system. By incorporating features from both the framework and the system, BigRoots is able to identify the reasons for stragglers from a broader spectrum.}
\item{We introduce statistical rules for different features to reduce false positive results of the root-cause analysis. We leverage \emph{edge detection} to prevent high resource utilization from normal task being considered as the root cause of stragglers. We derive lower bounds through empirical study for time related features to prevent insignificant features being considered as root cause.}
\item{We verify BigRoots with controlled experiments by injecting high resource utilization across different system components. We analyze different workloads in Hibench using BigRoots and discover unique root causes of stragglers for different workloads. The root-cause analysis of BigRoots is insightful to guide effective performance optimizations for programmers.}
\end{itemize}

The rest of the paper is organized as follows. Section~\ref{sec:background} presents research background of root-cause analysis of straggler in big data system. Section~\ref{sec:methodology} introduces our method that incorporates features from both the framework and the system for root-cause analysis. Section~\ref{sec:evaluation} evaluates the effectiveness of our approach by applying our approach to synthetic resource anomalies as well as representative workloads in Hibench. Section~\ref{sec:relatedwork} presents the related work in this field. Section~\ref{sec:conclusion} concludes our work.

%% file: background.tex
\section{Background}
\label{sec:background}
\subsection{Straggler}
Big data frameworks commonly divide a job into small tasks running simultaneously on many nodes. The duration of a job is determined by the slowest task. The deviation of task duration can be attributed to many reasons. First, different tasks deal with different data and the larger amount of data is, the slower the task will be (\textit{data skew}). Secondly, the data a task process may not reside in local node and retrieving data from remote node can cause significant delay especially when the network is congested. Moreover, different nodes may have different hardware configurations, which is another factor may cause difference in execution time. Especially for big data framework such as Hadoop and Spark, stragglers usually happen during the reduce phrase due to severe data skew. Straggler can be mitigated through many ways if the root causes of stragglers are known. If data skew is the root cause, we can change the key or split data into more partitions. If resource contention is the root cause, we can speculatively launch the task on the nodes with less load. Different researchers have different definitions of stragglers. Mantri~\cite{mantri} defines straggler as the task whose time to finish is longer than 1.5x the median task duration in its phrase. We adopt the same definition for straggler as Mantri in this paper. 
\subsection{Spark}
Spark is designed to speed up applications by reusing a working set of data across multiple parallel operations while maintaining the scalability and fault-tolerance properties of MapReduce. Spark introduces a data abstraction called resilient distributed datasets (RDDs). An RDD is a read-only collection of objects partitioned across a set of machines that can be rebuilt if a partition is lost. Spark normally splits a job into different stages. When a program invokes action or shuffle operation, it splits the procedure into different stages. Shuffle read/write data size at the end of each stage is an important factor for performance. Generally, Spark acts similarly to other big data frameworks except that it shows lower task launch overhead and breaks a job into more tasks which can reduce the impact of stragglers. We use Spark as the target framework to evaluate our approach for root-cause analysis of stragglers in this paper.


\subsection{Root Cause Analysis}
Root cause analysis is to identify the underlying reasons for stragglers. The result of such analysis can be used to optimize the performance of the big data application. For instance, if the root cause is due to shuffle too much data, we can optimize key partition or data layout to mitigate such straggler. Root cause analysis can be divided into task level and job level. Job level analysis only provides coarse-grained information about stragglers for a job or a group of jobs, whereas task level analysis provides detailed root cause for each straggler task. Existing works commonly use online analysis~\cite{mantri,dolly,performance_instrument} to perform task level analysis and offline analysis~\cite{cause_simple,straggler_impact} to perform job level analysis. The shortcoming of online analysis is that the available information is constrained and the accuracy of the analysis is quite limited. Previous works~\cite{straggler_impact, cause_simple, performance_instrument} also propose to identify the root causes through critical features such as high resource utilization. However such features could also appear with normal tasks and thus affects the accuracy of the analysis.

%% file: methodology.tex
\section{Root-cause Analysis with Effective Features}
\label{sec:methodology}
\subsection{Straggler Features}
BigRoots choose features that are commonly believed to have great impact on the performance of tasks including system features such as CPU, I/O, network utilization~\cite{cause_simple,hedera, tachyon,google_trace_analysis,titan,straggler_impact} and framework features including data locality, shuffle read/write bytes, bytes read, JVM garbage collection time, serialization time and deserialization time~\cite{mantri,dolly}.

\subsubsection{System Features}
We use several sampling tools to get system utilization features during task execution. For CPU utilization, we use \textit{Multiprocessor Statistics} (MPSTAT) to sample user CPU time every one second. CPU feature for a task is calculated as shown in Equation~\ref{eq:cpu}, where $t_0$ and $t_1$ is the start time and finish time of a task. Note that the user time and total time are averaged across all cores for multi-core CPUs. For disk utilization, we use \textit{Input/Output Statistics} (IOSTAT) to collect operating system storage input and output statistics every one second. Disk feature for a task is calculated as shown in Equation~\ref{eq:disk}, where $I/O\_time$ is the time system dealing with I/O requests. For network utilization, we use \textit{System Activity Report} (SAR) to sample network package exchange rate every one second. Network feature for a task is calculated as shown in Equation~\ref{eq:network}.

\begin{equation}
\label{eq:cpu}
F_{cpu}=\frac{1}{t_1-t_0}\sum_{t=t_0}^{t=t_1}{\frac{user\_time_t}{total\_time_t}}
\end{equation}

\begin{equation}
\label{eq:disk}
F_{disk}=\frac{1}{t_1-t_0}\sum_{t=t_0}^{t=t_1}{\frac{I/O\_time_t}{total\_time_t}}
\end{equation}

\begin{equation}
\label{eq:network}
F_{network}=\frac{1}{t_1-t_0}\sum_{t=t_0}^{t=t_1}{(Bytes\_sent_t+Bytes\_received_t)}
\end{equation}

\subsubsection{Framework Features}
Features extracted from Spark log files reflect internal root causes such as data skew, poor data locality and shuffle delay. Locality is a special feature incorporated in our approach that has several states in Spark as shown in Table~\ref{tab:locality}. Spark can record the corresponding wait time within each of these locality states by \emph{spark.locality.wait.*}. For instance, if current process has no available slot to process its local data, it enters into the \emph{spark.loaclity.wait.process} state and wait until timeout. After that, Spark will launch the task with degrading locality such as node level. According to our empirical study, locality feature is an important factor to determine the performance of Spark application. We use numerical value to represent the task locality feature as shown in Equation~\ref{eq:locality}. Other features used in our root-cause analysis are shown in Table~\ref{tab:other}.

\begin{equation}
\label{eq:locality}
\begin{aligned}
F_{locality}&=0, PROCESS\_LOCAL\\
&=1, NODE\_LOCAL\\
&=2, otherwise
\end{aligned}
\end{equation}

\begin{table}
\centering
\caption{Locality Features in Spark.}
\begin{tabular}{|c|m{4cm}|}
\hline
Locality&Meaning\\\hline
PROCESS LOCAL&data is in current process\\\hline
NODE LOCAL&data is in the same node\\\hline
RACK LOCAL&data is the node which is in the same rack as current node\\\hline
ANY&data in other nodes which not in the same rack\\\hline
NOPREF&there is no difference where data is, like reading from database\\\hline
\end{tabular}
\label{tab:locality}
\end{table}

\begin{table}
\centering
\caption{Other features extracted from Spark log files.}
\begin{tabular}{|c|m{1.5cm}|m{3cm}|@{}m{0pt}@{}}
\hline
Feature Name&Definition&Explanation\\\hline
$F_{read\_bytes}$&$R/R_{avg}$&Read bytes factor. B is read bytes in this task, $B_{avg}$ is the average read bytes in the stage. \\\hline
$F_{shuffle\_read\_bytes}$&$B/B_{avg}$&Shuffle read bytes factor.&\\[8pt]\hline
$F_{shuffle\_write\_bytes}$&$B/B_{avg}$&Shuffle write bytes factor.&\\[8pt]\hline
$F_{momery_bytes_spilled}$&$B/B_{avg}$&Bytes spilled into memory factor.\\\hline
$F_{disk_bytes_spilled}$&$B/B_{avg}$&Bytes spilled into disk factor.\\\hline
$F_{JVM_GC_time}$&$T/T_{task}$&Time spent in JVM garbage collection (GC) factor. T is GC time, $T_{task}$ is task duration.\\\hline
$F_{serialize_time}$&$T/T_{task}$&Time spent in result serialization factor.\\\hline
$F_{deserialize_time}$&$T/T_{task}$&Time spent in executor deserialization factor.\\\hline
\end{tabular}
\label{tab:other}
\end{table}

We define straggler task as whose duration is 1.5$\times$ larger than median task duration~\cite{mantri,straggler_impact,late} in the same stage. The idea of our root cause analysis is based on the following two observations: \textit{1)} If a feature is abnormal compared to tasks in other nodes, then this feature is highly possible the root cause of the straggler. \textit{2)} If a feature is abnormal compared to tasks in the same node, then this feature is highly possible the root cause of the straggler. The reason why we consider the intra-node and inter-node tasks separately is that the number of inter-node tasks is much larger than intra-node tasks. The importance of intra-node tasks would be underestimated if we consider intra-node and inter-node tasks inseparably.

\subsection{Root-cause Feature Identification}
\label{subsec:identification}
In BigRoots, the features we considered can be classified into four categories including discrete features, numerical features, resource features and time features. For a numerical feature, We consider it as the root cause when the conditions in Equation~\ref{eq:conditions} are satisfied, where $global\_quantile_{\lambda_{q}}$ is the $\lambda{q}$ quantile of feature values across all tasks, $F_{peer}$ is the feature of either inter-node tasks or inner-node tasks in the same stage, $\lambda_{p}$ is the threshold to adjust the sensitivity to straggler. The first condition aims to constrain the absolute value of the root cause. This is because even if the value of a straggler feature is larger than its peer task, it could still be normal variance and too weak to impact the duration of task. For time feature, we apply an additional rule, $F>0.2$. That is, the value of the feature must be 0.2 times larger than its peer task. 

For resource feature, we adopt another method to filter out the cases where high utilization is caused by the task itself. We call this method \emph{Edge Detection}. The idea is to monitor the system resource utilization for a small period before task begins and after task ends. If system resource utilization raises after task begins and drops after task ends, we will attribute the resource utilization to the job itself and thus the resource utilization should not be considered as the root cause. We filter out such resource feature if it satisfies the condition in Equation~\ref{eq:mean}, where $Mean^{head}_t$ and $Mean^{tail}_t$ is the mean resource utilization during time $t$ period before task begins and after task finishes. Discrete features represent the task locality. We consider locality as the root cause if the locality value is 2 and satisfies the condition in Equation~\ref{eq:sum}, where $num(normal\_task)$ is the number of normal tasks. This equation is based on the observation that straggler tasks usually access data remotely, whereas normal tasks mostly access local data.


\begin{equation}
\label{eq:conditions}
\begin{aligned}
&F>global\_quantile_{\lambda_{q}}\\
&F>mean(F_{peer})*\lambda_{p}\\
\end{aligned}
\end{equation}

\begin{equation}
\label{eq:mean}
\begin{aligned}
&Mean^{head}_t(F_{resource})>\lambda_eF_{resource}\\
&Mean^{tail}_t(F_{resource})>\lambda_eF_{resource}
\end{aligned}
\end{equation}

\begin{equation}
\label{eq:sum}
\begin{aligned}
sum(F_{locality}^{normal\_task})<num(normal\_task)/2
\end{aligned}
\end{equation}

\subsection{Putting All Together}
\begin{figure}[htbp]
\centering
\includegraphics[scale=0.5]{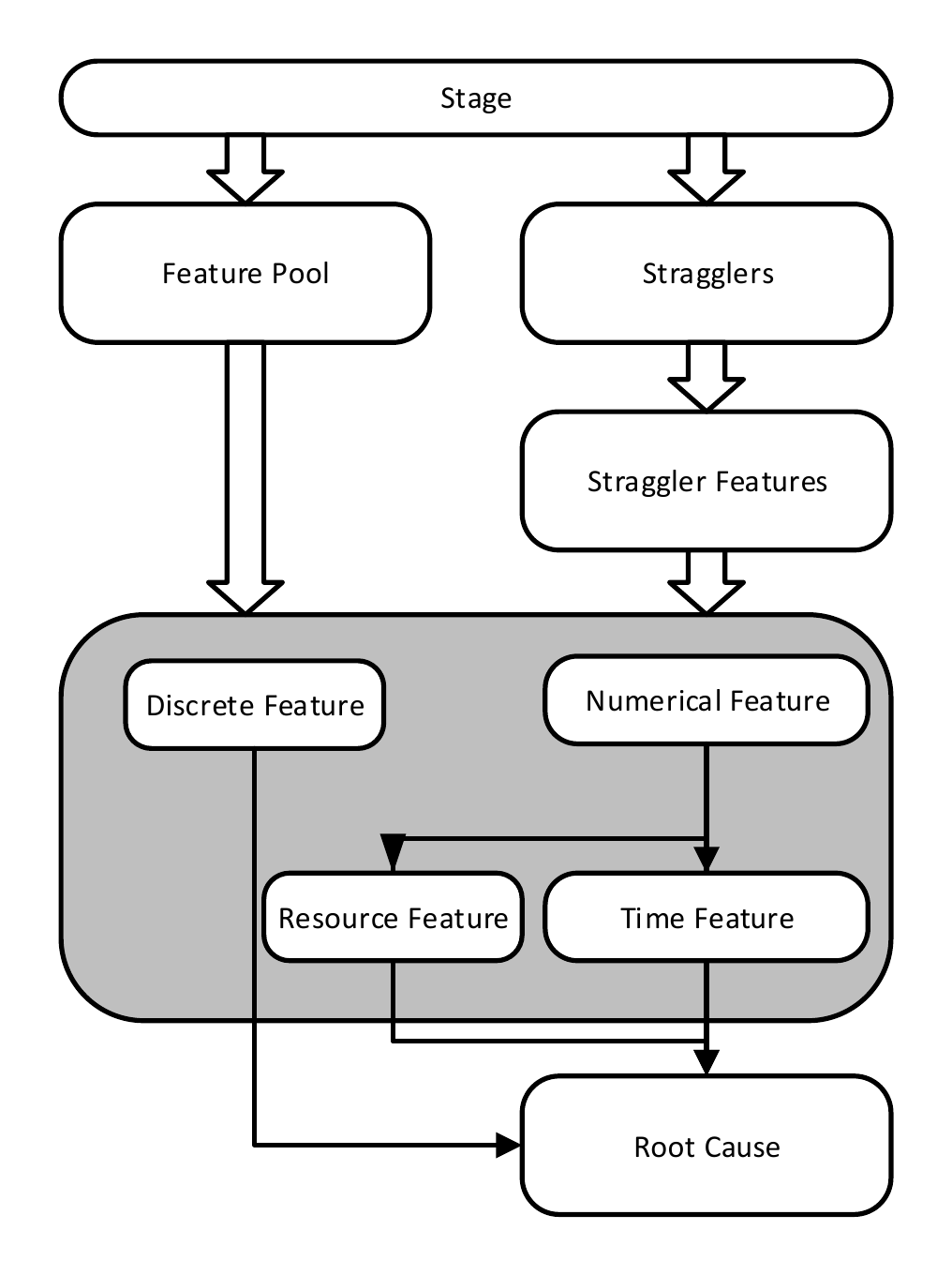}
\caption{The root-cause analysis workflow of BigRoots.}
\label{fig:overview}
\end{figure}

The root-cause analysis workflow of BigRoots is shown in Figure~\ref{fig:overview}. First, it collects features from both framework and system for all tasks within the same stage and forms a feature pool. Then stragglers whose execution time is 1.5$\times$ longer than median task are identified with straggler features collected. These straggler features are filtered again to identify the root cause features using the approach illustrated in section~\ref{subsec:identification}.

%% file: evaluation.tex
\section{Evaluation}
\label{sec:evaluation}
\subsection{Experiment Setup}
Our experiments are conducted from two aspects: \textit{1)} verify BigRoots by generating stragglers with controlled high resource utilization and test the accuracy of root-cause analysis with these stragglers. \textit{2)} analyze root causes of different stragglers on representative big data workloads in Hibench using BigRoots. Our experiment is conducted on a cluster of six servers. Each server is equipped with Intel Xeon E5-5620 which contains 16 core, 32KB L1 cache, 256KB L2 cache and 12MB L3 cache, 16G memories. The servers are connected with 1Gbps network. We use Spark v2.2.0 and HDFS v2.2.0. One server serves as master and other five servers serve as slaves. The Operating system is CentOS v6.5.

\begin{figure}[htbp]
\centering
\includegraphics[scale=0.55]{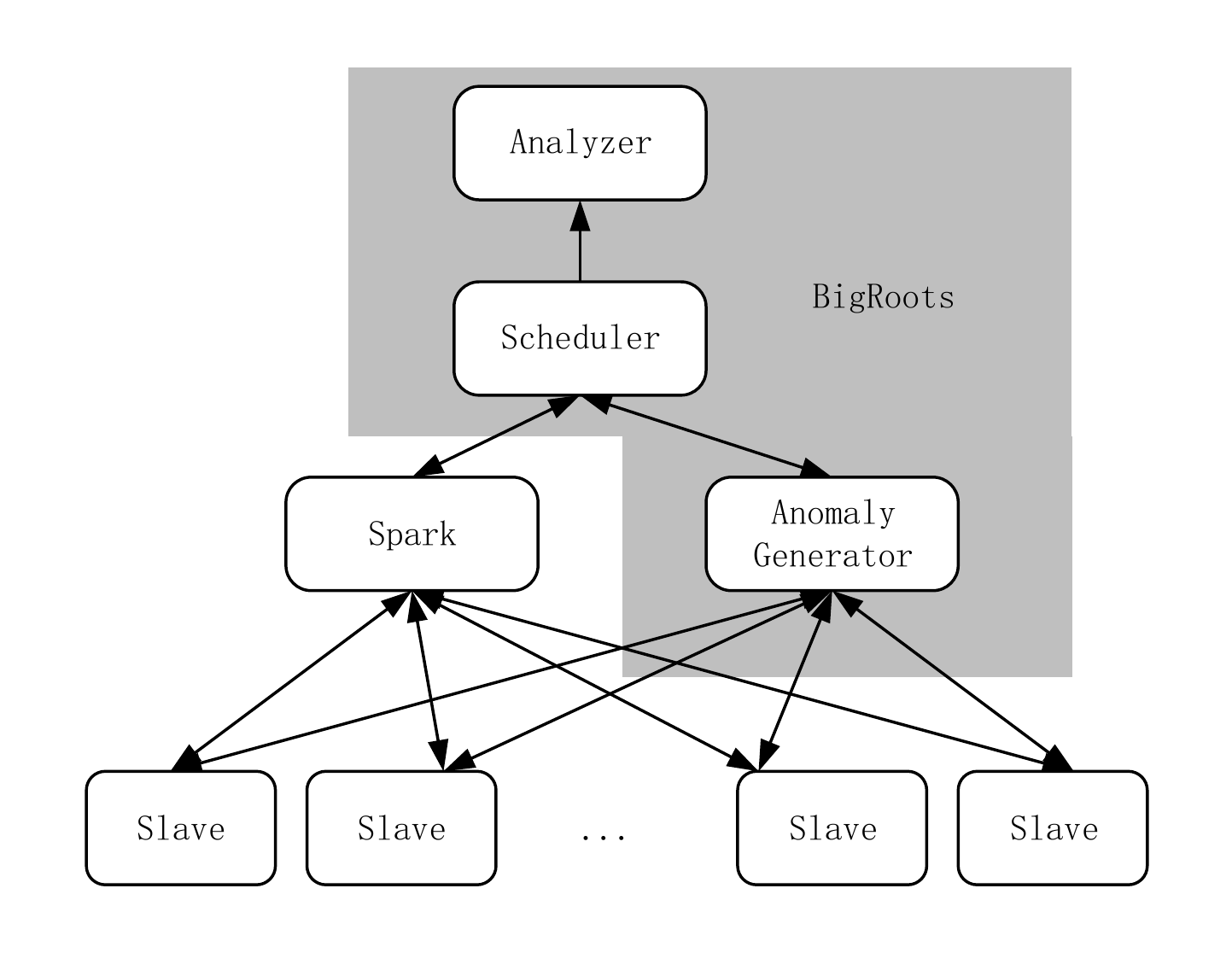}
\caption{The system design of implementing BigRoots on top of Spark.}
\label{framework}
\end{figure}

For evaluation, we implement BigRoots on top of Spark as shown in Figure \ref{framework}. BigRoots scheduler is in charge of dispatching Spark job and triggering Anomaly Generator (AG) to generate high CPU, I/O and network utilization. AG is in charge of launching resource hogging programs in slave nodes according to the decision from scheduler. Scheduler periodically collects information from Spark and AG log files, and send it to the analyzer. The analyzer extracts the features from the log information, organize them into task structure, and then performs the root-cause analysis. The design of the anomaly generator for each type of system resource is illustrated as follows.

\subsubsection{CPU Anomaly Generator}
CPU AG generates 1M random data and then performs power operation on each data in a loop to simulate computation intensive workload. We randomly dump one element of these data to disk in order to avoid compiler optimization and reduce disk utilization. Considered multi-core CPU, we launch eight processes to run CPU AG at the same time.
\subsubsection{I/O Anomaly Generator}
For I/O AG, we continuously write $10^8$ characters to disk in a loop to simulate I/O intensive workload. Similar to CPU AG, we launch eight processes at the same time.
\subsubsection{Network Anomaly Generator}
For network AG, we continuously send 512 characters to a remote TCP server and receive replies from the server to simulate workload exchanging massive data with remote host. The server is in the same LAN with the client to support large network flow. We also launch 8 processes at the same time.

For comparison, we implement Pearson Correlation Coefficient method (PCC), which has been used in root cause analysis by existing work~\cite{web_anomaly,web_correlation}. This PCC method provides a baseline for evaluating the accuracy of our approach. Equation~\ref{eq:pearson} shows that feature $F$ is identified as the root cause in PCC, where $\rho$ is Pearson Coefficient of current feature and task duration, $\lambda$ is coefficient.


\begin{equation}
\label{eq:pearson}
\begin{aligned}
&|\rho|>\lambda_{ca}\\
&F>quantile_{\lambda_{cq}}
\end{aligned}
\end{equation}


\subsection{Verification with Anomaly Generator}
\subsubsection{Injecting Single Anomaly}
\begin{figure}[htbp]
\centering
\includegraphics[scale=0.5]{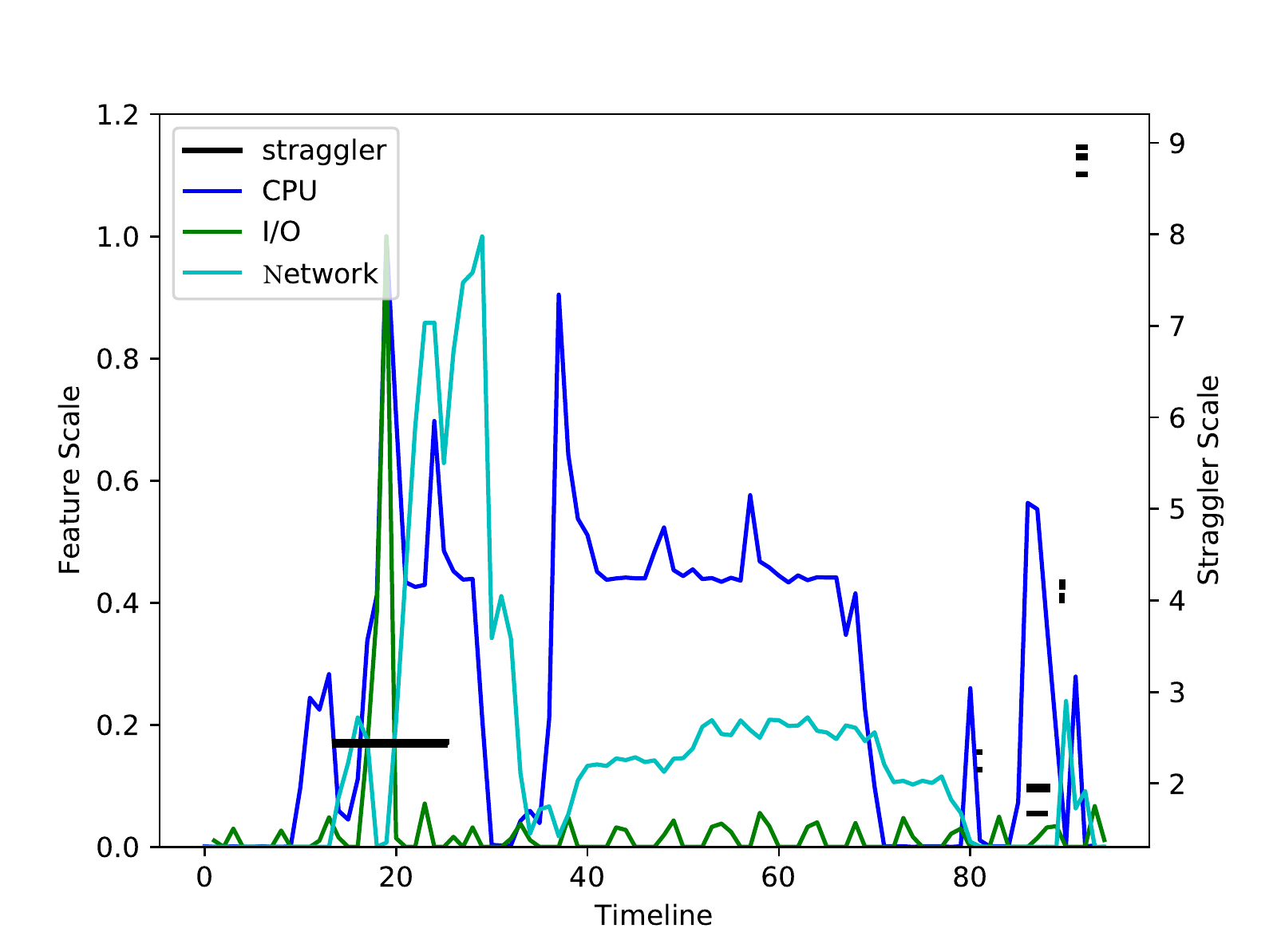}
\caption{Resource utilization and straggler scale when running NaiveBayes Classifier with no anamaly injection.}
\label{bayes-null}
\end{figure}

\begin{figure}[htbp]
\centering
\includegraphics[scale=0.5]{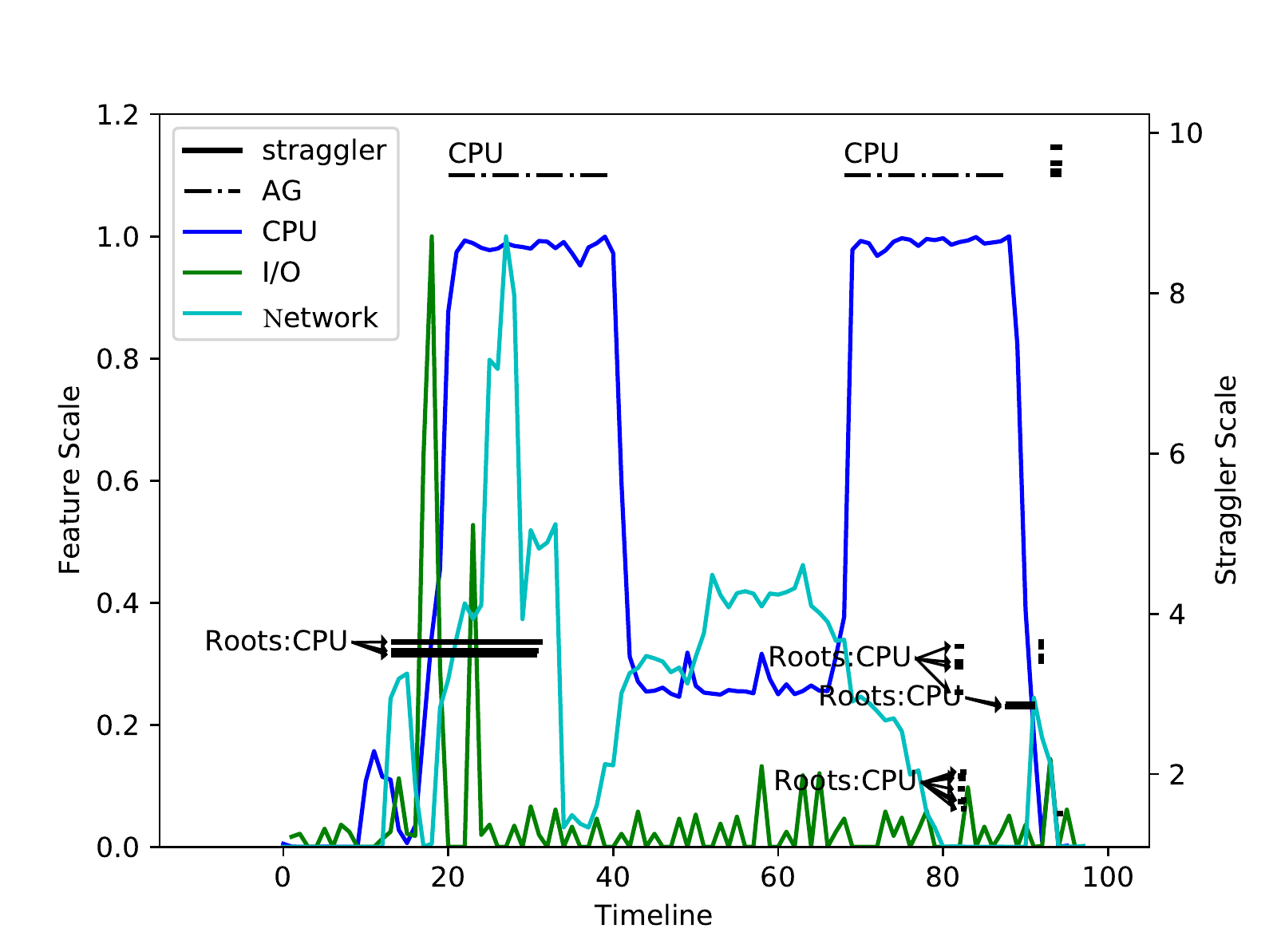}
\caption{Resource utilization and straggler scale when CPU AG generates high CPU contention.}
\label{bayes-cpu}
\end{figure}

\begin{figure}[htbp]
\centering
\includegraphics[scale=0.5]{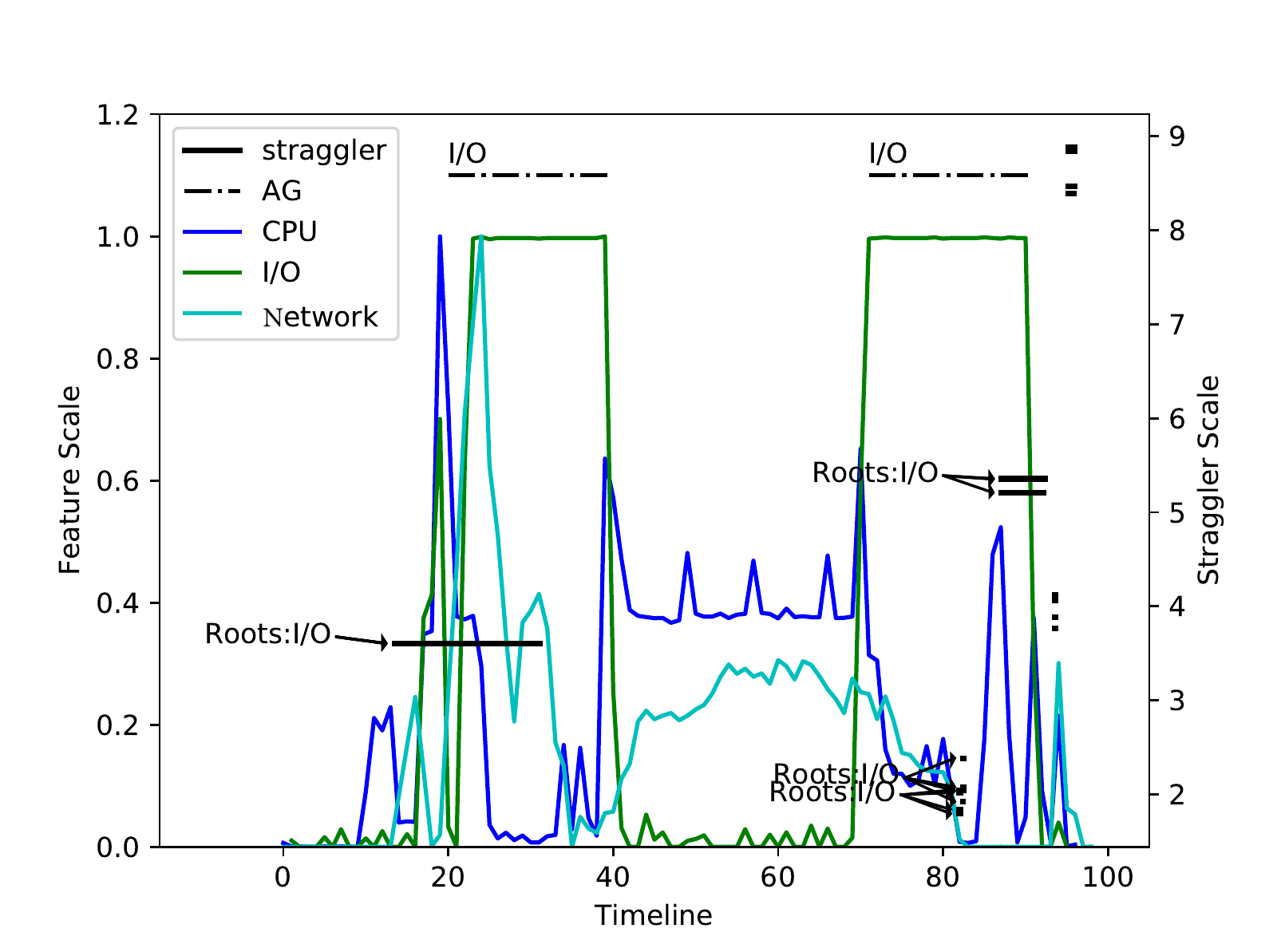}
\caption{Resource utilization and straggler scale when I/O AG generates high I/O contention.}
\label{bayes-io}
\end{figure}

\begin{figure}[htbp]
\centering
\includegraphics[scale=0.5]{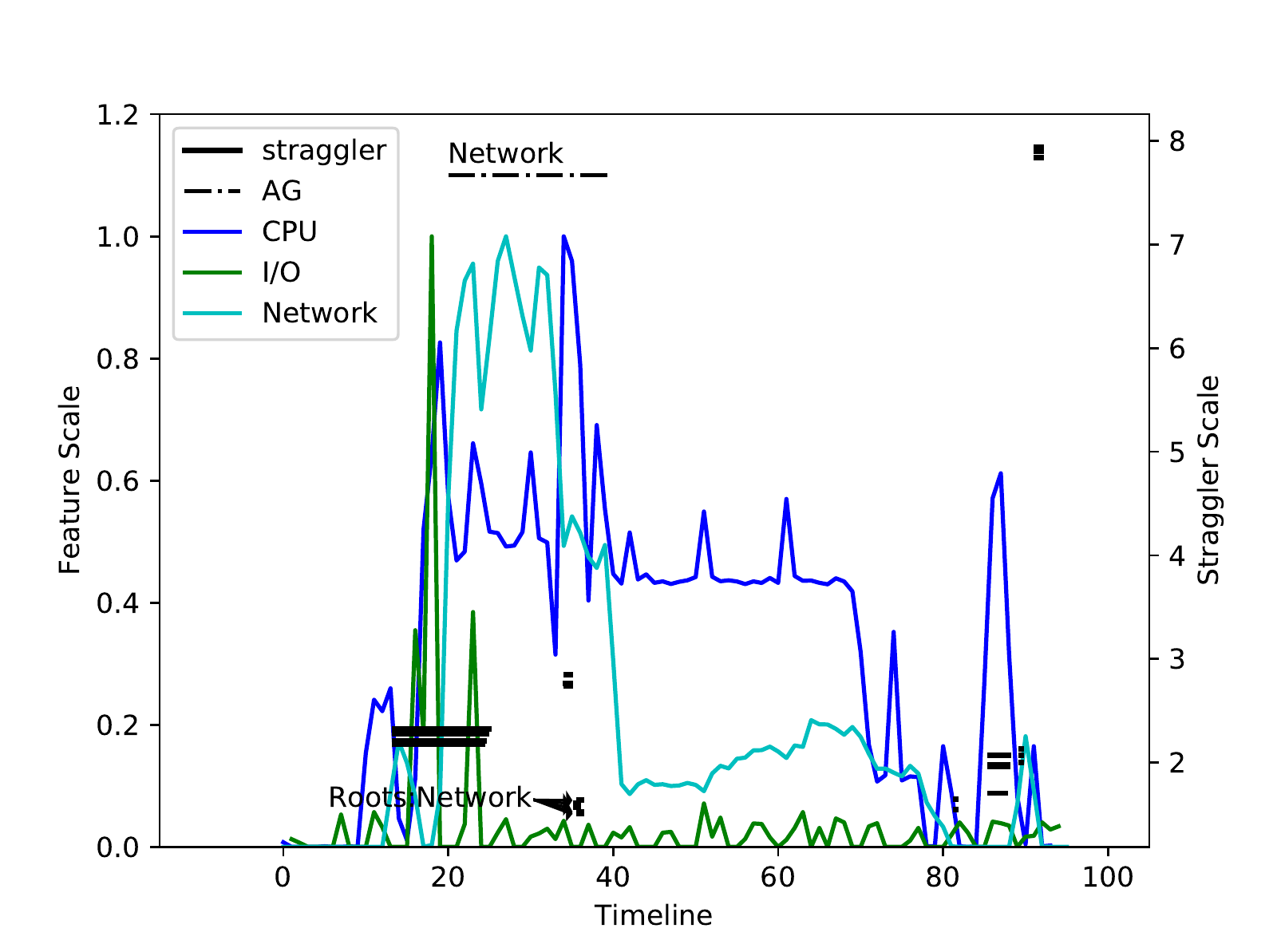}
\caption{Resource utilization and straggler scale when network AG generates high network contention.}
\label{bayes-net}
\end{figure}

We use NaiveBayes workload with large input (1 million pages and 100 classes) and apply different kinds of AG to verify BigRoots. We start AG in one salve node intermittently to simulate real cluster environment (resource utilization fluctuation). If a task's duration overlaps with AG Injecting period, we consider this task is influenced by the AG. And if the influence of an AG leads to a straggler, BigRoots should be able to identify the anomaly we injected as the root cause of straggler. Figure~\ref{bayes-null} to \ref{bayes-net} show the resource utilization and the scale of straggler as the workload of NaiveBayes Classifier executes. The $x$ axis is the timeline of the job. The $y$ axis on the left represents the utilization in percentage of the resource features. The $y$ axis on the right represents the scale of straggler that is calculated by dividing the duration of straggler by the duration of median task. The black lines in the figures indicate the straggler tasks identified with their root cause annotated. The black dash line indicates the time and duration different types of anomalies (generating high resource utilization) injected.

Figure \ref{bayes-cpu} shows the resource utilization and straggler scale when CPU AG generates high CPU contention. BigRoots is able to correctly attribute the root cause of stragglers impacted by AG to high CPU utilization. When we apply CPU AG, the straggler scale raise by 46\% from 2.43 to 3.55 with the timeline from 13s to 26s. High CPU utilization also causes more stragglers that do not exist in Figure~\ref{bayes-null} (from 82s to 85s) and BigRoots successfully identifies the root cause of these stragglers. Figure \ref{bayes-io} shows root cause BigRoots identified when I/O AG is injected. The straggler pattern is similar to that of CPU AG injected. However we can see I/O contention has more severe impact on straggler than CPU contention. The straggler scale increases by 2.6$\times$ around timeline 90s compared to Figure~\ref{bayes-null}. Nevertheless, BigRoots still accurately identifies the root cause of high I/O utilization. The network AG behaves quite differently as shown in Figure \ref{bayes-net}, which has a very little impact on the stragglers. It seems local area network (LAN) cannot become a performance bottleneck. This observation is in accordance with previous works~\cite{cause_simple,straggler_impact,performance_instrument} that concludes network congestion is hardly the root cause for stragglers. Therefore, only three stragglers in Figure~\ref{bayes-net} is annotated by BigRoots to the root cause of high network utilization.

The comparison between BigRoots and PCC is shown in Table \ref{PCC_comparison}. For PCC, we choose the best parameter setup through exhaustive search. We can see that BigRoots is much more accurate than PCC. PCC is more likely to generate false positive results. The reason is that straggler feature and task duration is not linearly correlated and features may correlate with each other. In addition, PCC is not very sensitive to outlier feature that usually shows great impact on straggler. In CPU anomaly injection experiment, although PCC identifies the same number of injected CPU anomalies as BigRoots, PCC gives a large number of false positive results that deteriorate the accuracy of root cause analysis. In I/O and network anomaly injection experiments, PCC identifies far less true positive results than BigRoots in addition to give many false positive results.

\begin{table}
\centering
\caption{Comparison between PCC and BigRoots. TP represents true positive, and FP represents false positive.}
\begin{tabular}{|c|c|c|c|c|}
\hline
\multirow{2}{*}{Experiment}&\multicolumn{2}{|c|}{BigRoots}&\multicolumn{2}{|c|}{PCC}\\\cline{2-5}
&TP&FP&TP&FP\\\hline
CPU AG&20&0&20&36\\\hline
I/O AG&16&0&9&29\\\hline
Network AG&5&4&5&21\\\hline
\end{tabular}
\label{PCC_comparison}
\end{table}

\begin{figure}[htbp]
\centering
\includegraphics[scale=0.5]{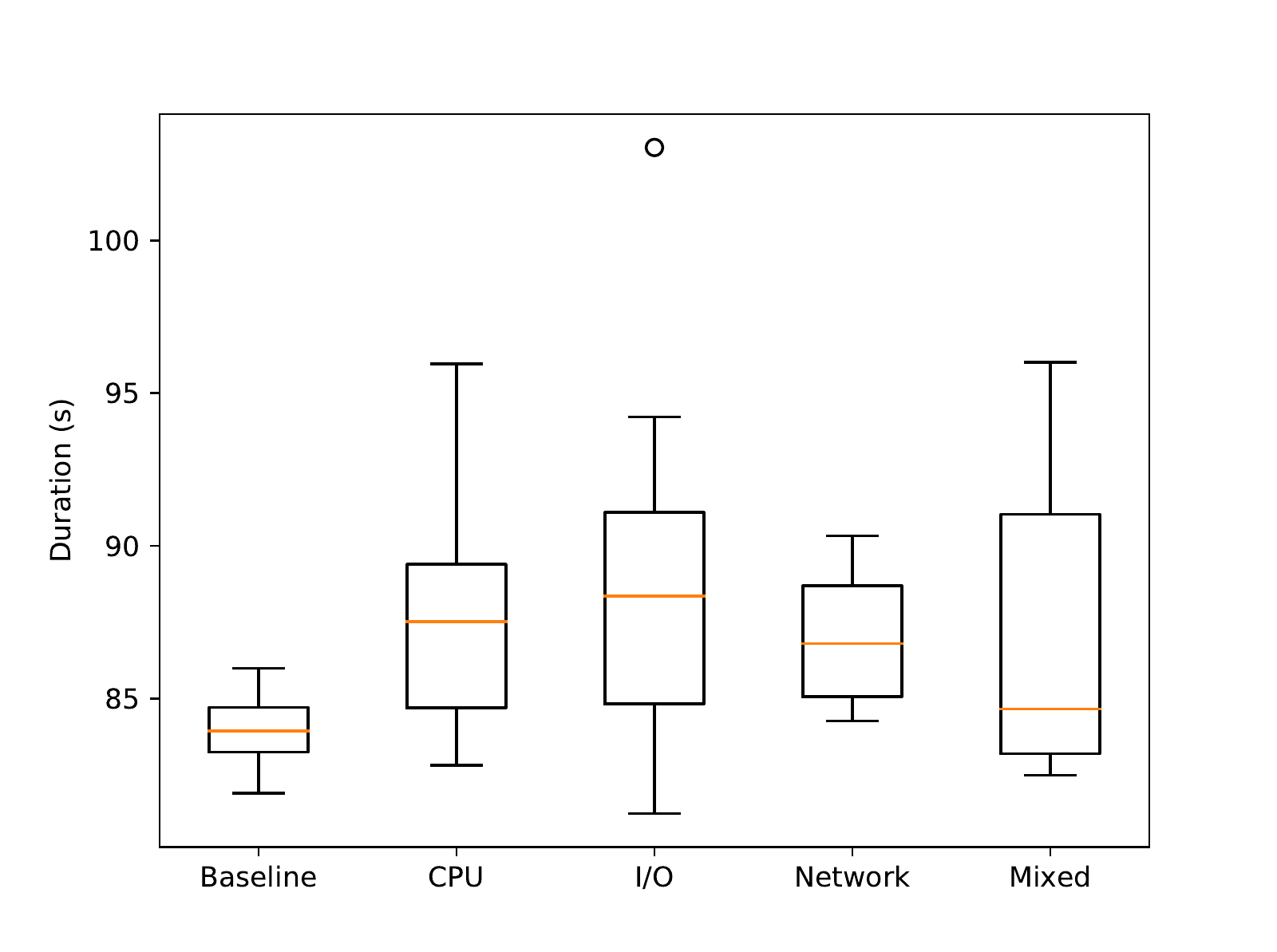}
\caption{Job duration when different AG is injected.}
\label{duration}
\end{figure}

We also evaluate the impact of separate AG as well as mixed AGs on job duration. With mixed AGs, all kinds of resource contention are randomly injected. The baseline shows the job duration with no AG applied. We repeat experiment 10 times and the result is shown in Figure \ref{duration}. We can see injecting I/O contention has the most impact on job duration which is contrary to the previous work~\cite{performance_instrument}. And network contention has the least impact on the job execution. The mean delay of CPU, I/O, network and mixed anomaly is 4.22\%, 5.86\%, 3.53\% and 4.02\% respectively. In general, none of these contentions causes severe performance degradation, which means the impact of resource contention on job execution is quite limited.

\subsubsection{Quantitative Analysis}
\begin{figure*}[htbp]
\centering
\includegraphics[scale=0.43]{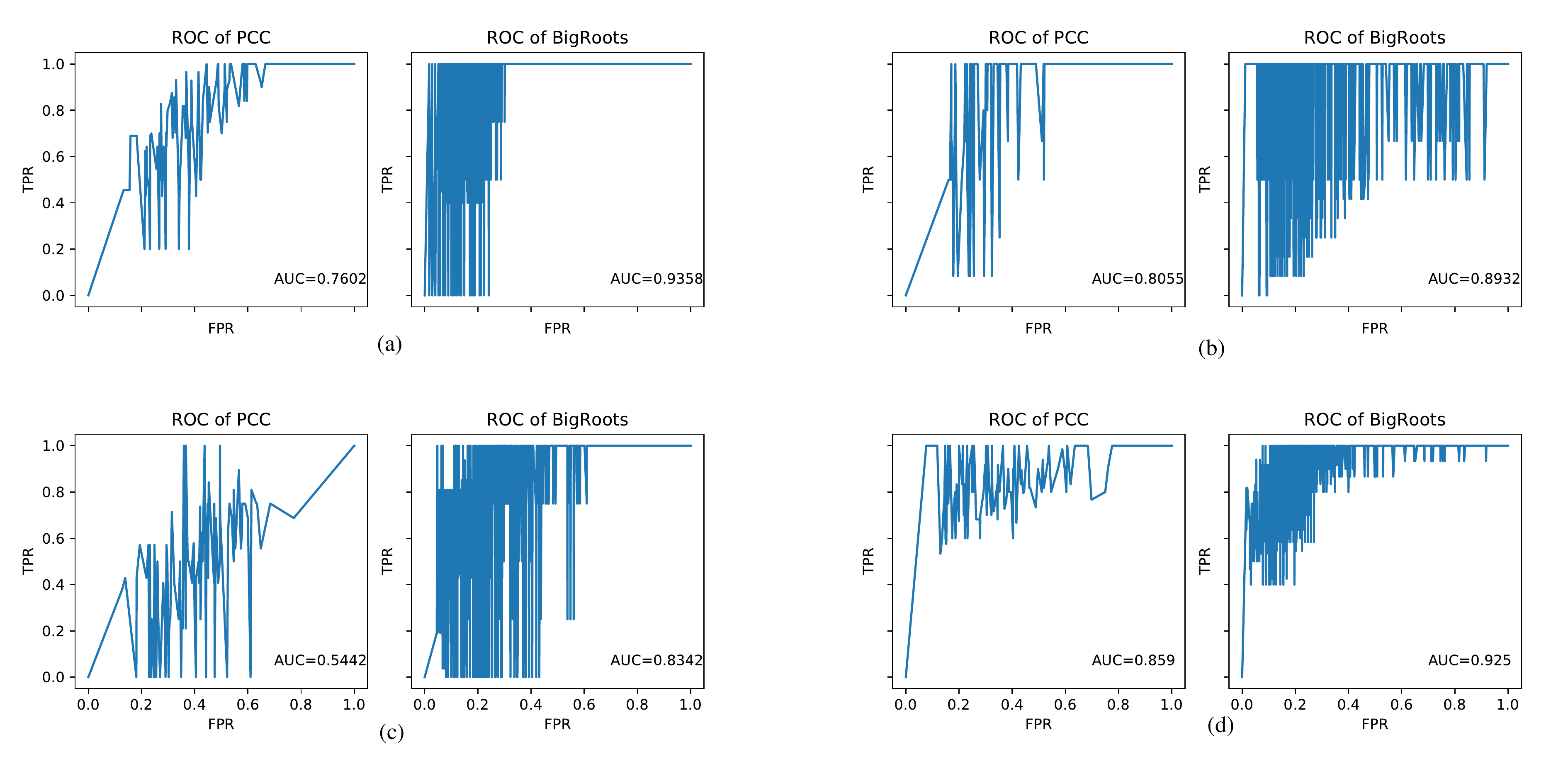}
\caption{The comparison of BigRoots and PCC. (a). ROC curve of BigRoots when CPU AG injected, (b). ROC curve of BigRoots and PCC when I/O AG injected, (c). ROC curve of BigRoots and PCC when network AG injected, (d). ROC curve of BigRoots and PCC when mixed AG injected.}
\label{fig:roc}
\end{figure*}

To understand the impact of threshold setup to the accuracy of root-cause analysis, we give \emph{receiver operating characteristic} (ROC) curve in Figure~\ref{fig:roc} under different thresholds. The $x$ axis of ROC curve is the \emph{false positive rate} (FPR) which represents the features that are incorrectly identified as the root cause by our approach. The $y$ axis of ROC curve is the \emph{true positive rate} (TPR) which represents the root-cause features that are correctly identified by our approach. The calculation of FPR and TPR is shown in Equation~\ref{eq:fprtpr}. TP (true positive) represents the feature affected by injected anomaly that is identified as the root cause by our approach. TN (true negative) represents the feature not affected by injected anomaly that is not identified as the root cause by our approach. FP (false positive) represents the feature not affected by injected anomaly that is incorrectly identified as the root cause by our approach. FN (false negative) represents the feature affected by injected anomaly that is not identified as the root cause by our approach.

\begin{equation}
\begin{aligned}
&FPR=FN/(FP+TN)\\
&TPR=TP(TP+FN)\\
&ACC=(TP+TN)/(TP+TN+FP+FN)
\end{aligned}
\label{eq:fprtpr}
\end{equation}

BigRoots use two thresholds during the root-cause analysis, \emph{quantile threshold} controls the difference between the current value and max value of the feature within the same stage, whereas \emph{median threshold} controls how much the current value is larger than the median value of the feature. Correlation method also uses two thresholds, \emph{Pearson threshold} controls minimum Pearson correlation coefficients and \emph{max threshold} controls how close feature value should be to max value within the same stage. We repeat each experiment 10 times to system noises. The fluctuation of ROC is caused by the joint influence of the two thresholds. BigRoots is better than correlation method if considering the \emph{area under curve} (AUC) for all experiments in Figure~\ref{fig:roc}. 

As shown in Figure~\ref{fig:roc}(a)-(c), when a single AG is injected, the AUC of BigRoots is larger than PCC by 23.10\%, 10.90\% and 53.29\% under CPU, I/O and network contention respectively. The ROC curve of BigRoots has more upper-left points which means it can better identify the root cause. The ROC curve of PCC is slightly above the diagonal line which means PCC is slightly better than a random guess. In Figure~\ref{fig:roc}(d) when mixed AGs are injected, the AUC of BigRoots is larger than PCC by 7.6\%. The reason for the similar accuracy under mixed resource contention between BigRoots and PCC is that the joint influence of different AGs leads to larger correlation coefficient and thus more detectable.

%
%
%

\subsubsection{Effect of Edge Detection}
\begin{figure*}[htbp]
\centering
\includegraphics[scale=0.55]{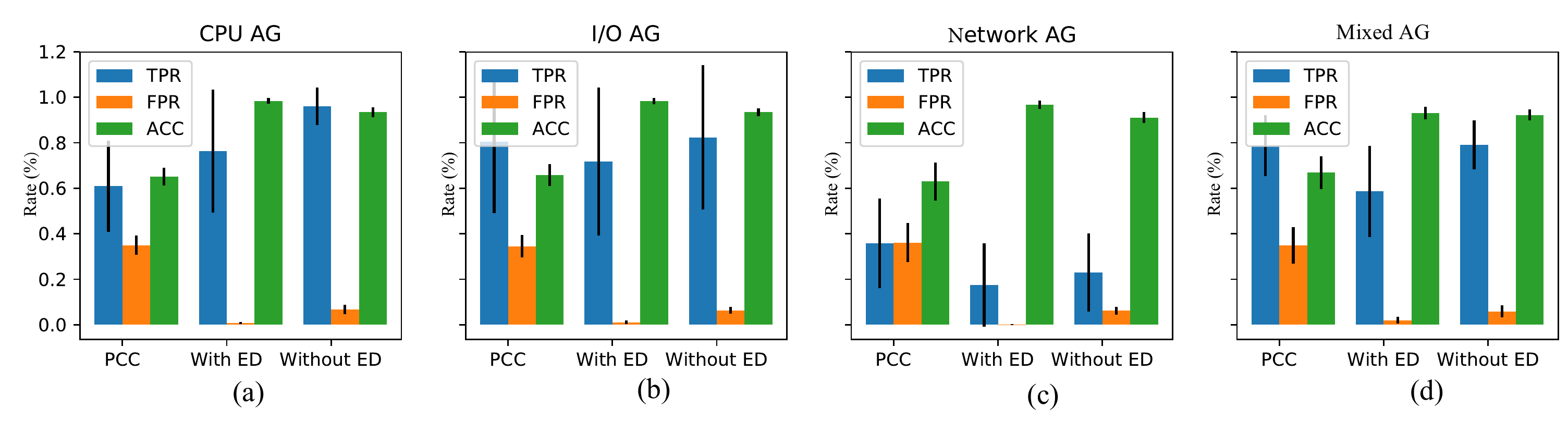}
\caption{Comparison of BigRoots with edge detection (root\_with\_edge), without edge detection (root\_no\_edge) and PCC when CPU, I/O, network, mixed AG injected.}
\label{fig:edge}
\end{figure*}

We demonstrate the effect of edge detection by comparing BigRoots with and without edge detection. The result is shown in Figure~\ref{fig:edge}. With edge detection, BigRoot decreases FPR by 85.71\%, 78.12\%, 100.00\%, 62.23\% and increases ACC by 0.88\%, 4.87\%, 6.53\%, 1.24\% when CPU, I/O, network, mixed AG is injected respectively. Edge detection improves BigRoots with much lower FPR and higher ACC, which indicates edge detection is effective in root cause identification under resource contention. Note that there are two thresholds for edge detection: edge width and filter threshold. Edge width controls the duration to monitor resource utilization before task starts and after task finishes. Filter threshold controls the sensitivity of the analysis towards resource utilization. Therefore, we can fine tune these two thresholds to further improve the effectiveness of edge detection. 

\subsubsection{Multiple Anomalies Across Nodes}
In order to evaluate BigRoots in real cluster scenario, we randomly start different AGs across different nodes for random periods. The time and duration when different AGs are injected across nodes is shown in Table~\ref{multi_nodes_AG}. The result of BigRoots and PCC is shown in Table~\ref{multi_nodes_comparison}. We can see that BigRoots has far smaller FRR (0.35\%) than PCC (16.25\%), which demonstrates BigRoots makes fewer mistakes by taking irrelevant feature as root cause. However, the TPR (recall) for BigRoots and PCC is 60.56\% and 66.19\% respectively, which is not promising enough. The reason is three-folds. First, the resource contention AG generates may not cause task delay. For instance, a task may read data only from local memory thus the network AG would not cause task delay. Secondly, the peer tasks may experience the same high resource utilization and both methods fail to identify the root cause in this case. Lastly, the duration of AG may only cover a small period of the straggler duration and is not the root cause for the straggler.

\begin{table}
\centering
\caption{The time and duration when different AGs are injected across different nodes.}
\begin{tabular}{|m{1.0cm}|c|c|}
\hline
Node & Time and Duration (s) & AG\\\hline
\multirow{2}{*}{Slave1}&0/10&CPU\\\cline{2-3}
&100/110&I/O\\\hline
\multirow{3}{*}{Slave2}&30/40&CPU\\\cline{2-3}
&63/73&CPU\\\cline{2-3}
&83/93&CPU\\\hline
Slave3&99/109&I/O\\\hline
\multirow{3}{*}{Slave4}&27/37&Network\\\cline{2-3}
&87/97&I/O\\\cline{2-3}
&112/122&Network\\\hline
\multirow{4}{*}{Slave5}&33/43&I/O\\\cline{2-3}
&53/63&CPU\\\cline{2-3}
&69/79&I/O\\\cline{2-3}
&100/110&CPU\\\hline
\end{tabular}
\label{multi_nodes_AG}
\end{table}

\begin{table}
\centering
\caption{The comparison of different methods to identify the root cause when multiple AGs are applied across multiple nodes.}
\begin{tabular}{|m{1.0cm}|c|c|c|c|c|c|c|}
\hline
Method &TP&TN&FP&FN&FPR (\%) &TPR (\%) &ACC (\%)\\\hline
BigRoots&43&282&1&28&0.35&60.56&91.81\\\hline
PCC&47&237&46&24&16.25&66.19&80.22\\\hline
\end{tabular}
\label{multi_nodes_comparison}
\end{table}

\subsection{Case Study with Hibench}
We use BigRoots analyzing different workloads in Hibench to find their root causes for stragglers. The result is shown in Table \ref{workloads}. It can be seen that machine learning applications are severely affected by data skew. Especially \textit{Kmeans} workload contains 49 stragglers caused by shuffle read. This is due to the disequilibrium of different clustering center in \textit{Kmeans}. Data skew happens when clustering centers are small and not evenly distributed. \textit{Naive Bayes} also suffers from data skew but there are fewer stragglers caused by data skew compared to \textit{Kmeans} (10 for \textit{Naive Bayes} and 49 for \textit{Kmeans}). This is in accordance with the fact that \textit{Naive Bayes} is vulnerable to data skew only when calculating the probability of different labels, but this calculation only occupies a small portion of the tasks. The keys of the remaining tasks are evenly distributed and thus data skew is unlikely to happen. \textit{Logistic Regression} and \textit{SVM} are both severely affected by data skew, therefore further performance optimization should focus on partitioning data more evenly. Specifically, both \textit{Logistic Regression} and \textit{SVM} are trained with \textit{Stochastic Gradient Descent} (SGD). It is highly possible the data skew is due to the SGD implementation in Spark. 

\textit{PCA} in Table~\ref{workloads} is the only machine learning workload not affected by data skew. However \textit{PCA} generates over 4000 stragglers and the root causes for most of them cannot be identified by BigRoots. The reason is three-fold. First, the thresholds in BigRoots are tuned during the AG injection experiments, which may not be optimal for other features. Secondly, the features used by BigRoots may not include all features that have an impact on task duration. Lastly, some stragglers can be affected by the joint effect of several less abnormal features. BigRoots fails to identify the root cause of such stragglers. On the other hand, the purpose of BigRoots is to identify the root cause of stragglers and provide guidance for performance optimization. However, these less abnormal features cannot provide much information for performance optimization. Micro, Graph, SQL and WebSearch workloads are mostly affected by resource contention and different workloads usually contend for different kinds of resources. For instance, \textit{Sort} workload mostly contends for I/O resource, \textit{Pagerank} and \textit{Nweight} workloads contend for CPU resource, in addition \textit{Nweight} workload also contends for network resource. Therefore, we can use faster disk to speedup \textit{Sort} and assign more CPU cores to speedup \textit{Nweight} and \textit{Pagerank}.

\begin{table}
\centering
\caption{Root cause analysis on Hibench workloads.}
\begin{tabular}{|m{1.2cm}|m{1.5cm}|m{2.2cm}|c|}
\hline
Domain &Workload &BigRoots Result& \# Stragglers\\\hline
\multirow{5}{*}{\parbox{1.2cm}{Machine Learning}}&Kmeans&I/O (4), CPU (4), shuffle\_read\_bytes (49)&151\\\cline{2-4}
&Naive Bayes&Shuffle\_read\_bytes (10)&155\\\cline{2-4}
&Logistic Regression&Bytes\_read(287), Network (1), I/O (4)&1084\\\cline{2-4}
&PCA&CPU (11), I/O (8), Network (8)&4107\\\cline{2-4}
&SVM&CPU (6), I/O (26), Network (167), Bytes\_read (1634)&4305\\\hline
\multirow{3}{*}{Micro}&Sort&I/O (4)&10\\\cline{2-4}
&Terasort&-&2\\\cline{2-4}
&Wordcount&-&9\\\hline
Graph&Nweight&CPU (7), Network (3)&97\\\hline
SQL&Aggregation&-&23\\\hline
WebSearch & Pagerank & CPU(4)&59\\\hline
\end{tabular}
\label{workloads}
\end{table}

\subsection{Overhead Analysis}
The overhead of BigRoots is mainly caused by resource sampling tools whose performance impact is negligible. The resource consumption of the sampling tools is shown in Table \ref{overhead}. These tools occupy less than 888KB memory and less than 1\% CPU.

\begin{table}
\centering
\caption{Resource consumption of the sampling tools.}
\begin{tabular}{|c|c|c|}
\hline
Sampling Tool & CPU Utilization (\%) & Memory Utilization (KB) \\\hline
mpstat&$0.5\pm0.2$&872\\\hline
iostat&$0.7\pm0.3$&864\\\hline
sar&$0.2\pm0.1$&888\\\hline
\end{tabular}
\label{overhead}
\end{table}

%

%% file: related_work.tex
\section{Related Work}
\label{sec:relatedwork}
Root-cause analysis is first introduced in big data to promote the efficiency of speculative execution of stragglers. G. Ananthanarayanan et al.~\cite{mantri} first introduce the idea of straggler root cause analysis to make speculative execution more efficient and implement it in Mantri. Their intuition is simple: if a task is slow because of data skew, then restart it on another machine won't make the task finish early. Mantri uses a top-down method to determine the root cause of a straggler. Mantri first checks whether this task's delay can be explained by processing too much data or reading too much data from the network. Their experiments show that in 40\% of the phases all tasks can be explained by this factor. When a straggler cannot be explained by data skew Mantri will check crossrack traffic. Lastly, Mantri attributes stragglers to bad and busy machines.

K. Ousterhout et al.~\cite{performance_instrument} improve root-cause analysis accuracy by introducing an instrumentation method to make sense of the influence of factors. They use two SQL benchmarks and a production workload analysing Spark framework and draw the conclusion that CPU is the main cause of straggler, not I/O. Their analysis is based on the block time of impact factors such as network, disk, etc.
They assume that a straggler is caused by X if it would not have been considered as straggler if a straggler had X taken zero time for all of the tasks in the stage. They finally attribute more than 60\% of stragglers to scheduler delay, HDFS disk read time, shuffle read time, shuffle write time, Java’s garbage collection. Their root-cause identification method needs instrumentation and causes overhead. Besides, their method can only attribute to factors whose block time can be measured.

Some works try to identify the job level root cause of stragglers using offline log analysis. X. Ouyang et al.~\cite{cause_simple} use correlation analysis in a 20-day period large-scale production Cloud datacenter composed of 5000 servers and millions of tasks. They simply count the number of stragglers where CPU utilization larger than 80\%, disk usage larger than 80\% and read-write request handling time of file system longer than 400ms. They observed 59\%, 42\% and 34.3\% of stragglers occur under the presence of high server CPU, disk overloading and slow request handling respectively.

Human analysis is used to the identify root cause of stragglers by~\cite{straggler_impact,tsar,nagios}. Tsar~\cite{tsar} and Nagios~\cite{nagios} monitor system metrics at a specific time interval and alerts potential atypical system behavior to technical staff. They simply throw the problems to human and fail to work properly when cluster grows bigger and hundreds of jobs simultaneously running. P. Garraghan et al.~\cite{straggler_impact} pick extreme straggler from historical data using Degree of Straggler comprising task execution time and input size which rules out stragglers caused by data skew and perform artificial analysis. They also attribute straggler to high resource utilization such as CPU, disk, unhandled operational access request, network whose occurrence frequency is 30\%, 23\%, 23\%, 13\% respectively.

In cloud computing area, previous works have proposed statistical and machine learning methods to analysis the root cause of anomalies.~\cite{web_bayesian} use a Bayesian classifier to identify the root cause of web anomalies.~\cite{web_unsupervised} proposes an unsupervised machine learning to identify anomalies. Statistical methods are more common in root-cause analysis.~\cite{web_anomaly,web_correlation} propose a root cause analysis method using Pearson coefficient of correlation between aggregated workload, latency and system metrics.~\cite{icdcs_holistic_analysis} proposes an anomaly detection method for bottleneck identification which monitors the applications, operators, and infrastructure level operations.~\cite{icdcs_utilization_monitor} proposes an method for millisecond-level bottleneck identification by comparing resource utilization among distributed nodes. H. Jayathilaka et al.~\cite{web_rootcause} propose a synthetic statistical root-cause analysis method to study the performance anomalies in web applications deployed in Platform-as-a-Service (PaaS) clouds. They monitor the relative importance, change in relative importance, high quantiles, tail end values and select candidates. They used fault injection to verify their algorithm as we do in this paper.
Big data is similar to web service in that they are all distributed systems. The root cause analysis method used in web service is also valuable for big data bottleneck analysis except that big data should focus more on features across different tasks rather than different layers in web service.

%% file: conclusion.tex
\section{Conclusion}
\label{sec:conclusion}
In this paper, we propose a new method for root cause analysis of stragglers in big data system based on effective features from both framework and system. We have demonstrated the effectiveness of BigRoots by injecting resource contention across different system components compared to PCC. BigRoots successfully identifies the root cause of the stragglers affected by the resource anomaly injected. We also improve the accuracy of BigRoot by incorporating the method of edge detection, which significantly reduces false positive results on resource features. As case study, we evaluate BigRoots by analyzing different workloads in Hibench. Our approach is effective to identify the root causes of representative workloads and provide useful insights for further performance optimization. 

For the future work, we would like to extend BigRoots to incorporate more features in order to fully understand the root cause of the straggler. In addition, we would like to consider the correlation between different features, which helps us to identify the complicated root cause where features are not independent of each other. For instance, poor locality may be correlated with high network utilization, which forces the tasks to fetch data from remote nodes.



%% file: acknowledgement.tex
\section*{Acknowledgement}
This work is supported by National Key Research and Development Program of China (Grant No. 2016YFB1000304) and National Natural Science Foundation of China (Grant No. 61502019). Hailong Yang is the corresponding author.

%% file: bare_conf.bbl
\begin{thebibliography}{10}
\providecommand{\url}[1]{#1}
\csname url@samestyle\endcsname
\providecommand{\newblock}{\relax}
\providecommand{\bibinfo}[2]{#2}
\providecommand{\BIBentrySTDinterwordspacing}{\spaceskip=0pt\relax}
\providecommand{\BIBentryALTinterwordstretchfactor}{4}
\providecommand{\BIBentryALTinterwordspacing}{\spaceskip=\fontdimen2\font plus
\BIBentryALTinterwordstretchfactor\fontdimen3\font minus
  \fontdimen4\font\relax}
\providecommand{\BIBforeignlanguage}[2]{{%
\expandafter\ifx\csname l@#1\endcsname\relax
\typeout{** WARNING: IEEEtran.bst: No hyphenation pattern has been}%
\typeout{** loaded for the language `#1'. Using the pattern for}%
\typeout{** the default language instead.}%
\else
\language=\csname l@#1\endcsname
\fi
#2}}
\providecommand{\BIBdecl}{\relax}
\BIBdecl

\bibitem{mapreduce}
J.~Dean and S.~Ghemawat, ``Mapreduce: simplified data processing on large
  clusters,'' in \emph{Conference on Symposium on Opearting Systems Design \&
  Implementation}, 2008, pp. 10--10.

\bibitem{dryad}
M.~Isard, M.~Budiu, Y.~Yu, A.~Birrell, and D.~Fetterly, ``Dryad: distributed
  data-parallel programs from sequential building blocks,'' in \emph{ACM SIGOPS
  operating systems review}, vol.~41, no.~3.\hskip 1em plus 0.5em minus
  0.4em\relax ACM, 2007, pp. 59--72.

\bibitem{spark}
M.~Zaharia, M.~Chowdhury, M.~J. Franklin, S.~Shenker, and I.~Stoica, ``Spark:
  Cluster computing with working sets.'' \emph{HotCloud}, vol.~10, no. 10-10,
  p.~95, 2010.

\bibitem{mantri}
G.~Ananthanarayanan, S.~Kandula, A.~G. Greenberg, I.~Stoica, Y.~Lu, B.~Saha,
  and E.~Harris, ``Reining in the outliers in map-reduce clusters using
  mantri.'' in \emph{OSDI}, vol.~10, no.~1, 2010, p.~24.

\bibitem{straggler_google}
J.~Dean and L.~A. Barroso, ``The tail at scale,'' \emph{Communications of the
  ACM}, vol.~56, no.~2, pp. 74--80, 2013.

\bibitem{straggler_impact}
P.~Garraghan, X.~Ouyang, R.~Yang, D.~McKee, and J.~Xu, ``Straggler root-cause
  and impact analysis for massive-scale virtualized cloud datacenters,''
  \emph{IEEE Transactions on Services Computing}, 2017.

\bibitem{google_estimation}
J.~Dean and S.~Ghemawat, ``Mapreduce: simplified data processing on large
  clusters,'' \emph{Communications of the ACM}, vol.~51, no.~1, pp. 107--113,
  2008.

\bibitem{late}
M.~Zaharia, A.~Konwinski, A.~D. Joseph, R.~H. Katz, and I.~Stoica, ``Improving
  mapreduce performance in heterogeneous environments.'' in \emph{Osdi},
  vol.~8, no.~4, 2008, p.~7.

\bibitem{dolly}
G.~Ananthanarayanan, A.~Ghodsi, S.~Shenker, and I.~Stoica, ``Effective
  straggler mitigation: Attack of the clones.'' in \emph{NSDI}, vol.~13, 2013,
  pp. 185--198.

\bibitem{speculation_waste}
E.~Bortnikov, A.~Frank, E.~Hillel, and S.~Rao, ``Predicting execution
  bottlenecks in map-reduce clusters,'' in \emph{Proceedings of the 4th USENIX
  conference on Hot Topics in Cloud Ccomputing}.\hskip 1em plus 0.5em minus
  0.4em\relax USENIX Association, 2012, pp. 18--18.

\bibitem{cause_simple}
X.~Ouyang, P.~Garraghan, R.~Yang, P.~Townend, and J.~Xu, ``Reducing late-timing
  failure at scale: straggler root-cause analysis in cloud datacenters,'' in
  \emph{Fast Abstracts in the 46th Annual IEEE/IFIP International Conference on
  Dependable Systems and Networks}.\hskip 1em plus 0.5em minus 0.4em\relax DSN,
  2016.

\bibitem{performance_instrument}
K.~Ousterhout, R.~Rasti, S.~Ratnasamy, S.~Shenker, B.-G. Chun, and V.~ICSI,
  ``Making sense of performance in data analytics frameworks.'' in \emph{NSDI},
  vol.~15, 2015, pp. 293--307.

\bibitem{hedera}
M.~Al-Fares, S.~Radhakrishnan, B.~Raghavan, N.~Huang, and A.~Vahdat, ``Hedera:
  Dynamic flow scheduling for data center networks.'' in \emph{NSDI}, vol.~10,
  2010, pp. 19--19.

\bibitem{tachyon}
H.~Li, A.~Ghodsi, M.~Zaharia, S.~Shenker, and I.~Stoica, ``Tachyon: Reliable,
  memory speed storage for cluster computing frameworks,'' in \emph{Proceedings
  of the ACM Symposium on Cloud Computing}.\hskip 1em plus 0.5em minus
  0.4em\relax ACM, 2014, pp. 1--15.

\bibitem{google_trace_analysis}
C.~Reiss, A.~Tumanov, G.~R. Ganger, R.~H. Katz, and M.~A. Kozuch,
  ``Heterogeneity and dynamicity of clouds at scale: Google trace analysis,''
  in \emph{Proceedings of the Third ACM Symposium on Cloud Computing}.\hskip
  1em plus 0.5em minus 0.4em\relax ACM, 2012, p.~7.

\bibitem{titan}
J.~Shi, Y.~Qiu, U.~F. Minhas, L.~Jiao, C.~Wang, B.~Reinwald, and F.~{\"O}zcan,
  ``Clash of the titans: Mapreduce vs. spark for large scale data analytics,''
  \emph{Proceedings of the VLDB Endowment}, vol.~8, no.~13, pp. 2110--2121,
  2015.

\bibitem{web_anomaly}
J.~P. Magalhaes and L.~M. Silva, ``Detection of performance anomalies in
  web-based applications,'' in \emph{Network Computing and Applications (NCA),
  2010 9th IEEE International Symposium on}.\hskip 1em plus 0.5em minus
  0.4em\relax IEEE, 2010, pp. 60--67.

\bibitem{web_correlation}
J.~P. Magalh{\~a}es and L.~M. Silva, ``Root-cause analysis of performance
  anomalies in web-based applications,'' in \emph{Proceedings of the 2011 ACM
  Symposium on Applied Computing}.\hskip 1em plus 0.5em minus 0.4em\relax ACM,
  2011, pp. 209--216.

\bibitem{tsar}
T.~tools, \url{https://github.com/alibaba/tsar/}.

\bibitem{nagios}
Nagios, \url{https://www.nagios.org/}.

\bibitem{web_bayesian}
X.~Gu and H.~Wang, ``Online anomaly prediction for robust cluster systems,'' in
  \emph{Data Engineering, 2009. ICDE'09. IEEE 25th International Conference
  on}.\hskip 1em plus 0.5em minus 0.4em\relax IEEE, 2009, pp. 1000--1011.

\bibitem{web_unsupervised}
D.~J. Dean, H.~Nguyen, and X.~Gu, ``Ubl: Unsupervised behavior learning for
  predicting performance anomalies in virtualized cloud systems,'' in
  \emph{Proceedings of the 9th international conference on Autonomic
  computing}.\hskip 1em plus 0.5em minus 0.4em\relax ACM, 2012, pp. 191--200.

\bibitem{icdcs_holistic_analysis}
A.~Arefin, V.~K. Singh, G.~Jiang, Y.~Zhang, and C.~Lumezanu, ``Diagnosing data
  center behavior flow by flow,'' in \emph{Distributed Computing Systems
  (ICDCS), 2013 IEEE 33rd International Conference on}.\hskip 1em plus 0.5em
  minus 0.4em\relax IEEE, 2013, pp. 11--20.

\bibitem{icdcs_utilization_monitor}
C.-A. Lai, J.~Kimball, T.~Zhu, Q.~Wang, and C.~Pu, ``milliscope: a fine-grained
  monitoring framework for performance debugging of n-tier web services,'' in
  \emph{Distributed Computing Systems (ICDCS), 2017 IEEE 37th International
  Conference on}.\hskip 1em plus 0.5em minus 0.4em\relax IEEE, 2017, pp.
  92--102.

\bibitem{web_rootcause}
D.~J. Dean, H.~Nguyen, and X.~Gu, ``Ubl: Unsupervised behavior learning for
  predicting performance anomalies in virtualized cloud systems,'' in
  \emph{Proceedings of the 9th international conference on Autonomic
  computing}.\hskip 1em plus 0.5em minus 0.4em\relax ACM, 2012, pp. 191--200.

\end{thebibliography}
